\documentclass[a4paper,showpacs,prb,twocolumn]{revtex4}
\usepackage{amsfonts}
\usepackage{graphicx}
\usepackage{amsmath}
\usepackage{amssymb}
\usepackage{latexsym}
\begin{document}

\title{First-principles study of electron transport in few-electron open quantum dots by the Hartree-Fock approach}
\author{S. Ihnatsenka}
\affiliation{Department of Physics, Simon Fraser University, Burnaby, British Columbia, Canada V5A 1S6}
\date{\today }

\begin{abstract}
Electron transport properties of few-electron open quantum dots within the spin-restricted Hartree-Fock approximation are studied. The self-consistent numerical calculations were performed for a whole device, including the semi-infinite leads, without employing any phenomenological or adjustable parameters. Inclusion of the non-local Fock potential brings qualitatively new physics in comparison to the Hartree approach: electron screening decreases, resonant energy levels become less pinned to the Fermi energy and clearly correlate with conductance peaks. When coupling between the dot and leads decreases the number of electrons inside the dot becomes quantized and the model predicts the Coulomb blockade of electron transport. This is confirmed by comparison with the master equation approach for an equivalent quantum dot.
\end{abstract}

 \maketitle

\section{Introduction}

Electron transport in mesoscopic structures is usually well described within the mean-field approaches that include electron-electron interaction at various levels of sophistication.\cite{Datta_book, dotPRL, Henrickson, Beenakker91, Jovanovic, Palacios05, Averin91, Meir91, Wacker, qpcnegf, Ihnatsenka, agreeOthers, QPC, Toher, Koentopp, Evers, Stefanucci04, Kurth10, Darancet07, opendot, Hirose, Feretti05, Indlekofer05, ABinterfer} The simplest model accounting for classical Coulomb repulsion between charged particles is the Hartree approach. While being the simplest it is able to capture many transport phenomena that are inaccessible for the single-particle approaches or the Thomas-Fermi model. For example, conductance statistics in open quantum dots at low temperatures was revealed to be governed by the pinning effect of resonant energy levels.\cite{dotPRL} However many phenomena observed experimentally demand further improvements of theoretical study such as inclusion of spin and of the exchange interaction. The hallmark example is the Coulomb blockade effect in quantum dots.\cite{Kastner} It is usually described by a phenomenological approach involving a set of adjustable parameters along with completely disregarding the dot geometry.\cite{Henrickson, Beenakker91, Jovanovic, Palacios05, Averin91, Meir91} As a result it is difficult to relate the employed parameter sets to the physical processes they represent and to the device topology studied. A model that accounts for both the dot and lead geometry and treats the whole device on the same footing in the Coulomb blockade regime as well as accesses direct quantitative description of Coulomb blockade in dots of general geometry has not been accomplished yet. Such a model would allow one to capture the microscopic physics of electron transport in the quantum dots.

Attempts to address Coulomb blockade, at least at a model level, are made within time-dependent density-functional theory (DFT).\cite{Stefanucci04, Kurth10} Electron transport is considered to be intrinsically non-equilibrium that evolve in time and Coulomb blockade occurs in the long-time limit as a periodic sequence of charging and discharging events. Electron tunneling through the interacting many-body system can also be considered within GW-approximation in the framework of non-equilibrium Green's functions (NEGF).\cite{Darancet07} By introducing quasi-particle electronic structure it brings substantial improvements in comparison to standard NEGF and agreement with experiment becomes better.\cite{Darancet07} However, whether time-dependence is necessary for steady-state regime and whether electronic correlations along are responsible for Coulomb blockade are not clear.

This theoretical study shows that inclusion of the exchange interaction in the Hartree-Fock approximation to the self-consistent solution of the Shr\"{o}dinger equation does allow one to capture the Coulomb blockade effect of electron transport. There is no need in the correlation effects, at least, for qualitative agreement. The model is implemented within the Green's function formalism that is well suited to the description of structures of arbitrary geometries.\cite{Datta_book, dotPRL, qpcnegf, Ihnatsenka, QPC, opendot, ABinterfer} It was previously tested (without the Fock term) on different low-dimensional structures and the results obtained were in good agreement with experimental data and other theoretical models.\cite{dotPRL, qpcnegf, Ihnatsenka} An attempt to account for exchange interaction was previously done using DFT in the local density approximation. The conductance however was found to be untrustworthy: It did not reproduce the 0.7 anomaly measured in almost every experiment on quantum point contacts.\cite{QPC} This is because the local density approximation of DFT lacks the derivative discontinuity and cannot fundamentally address the transport regime when quantization of electron charge occurs.\cite{Toher, Koentopp, Evers, Kurth10, QPC} Another approach that accounts for the exchange interaction \textit{exactly} is the Hartree-Fock one. It relies essentially on the density matrix that takes into account the phase relations between different electron states. Accounting for the exchange interaction by the density matrix brings qualitatively new physics in comparison to the Hartree approach. In the regime of a single channel open for transmission, the resonant levels in few-electron quantum dots become ordered at the Fermi energy and conductance oscillations narrow. Both of these phenomena are indicative of suppressed screening. As potential barriers are imposed between the dot and leads, the number of electrons in the dot becomes quantized and the conductance shows well defined peaks. To identify the transport regime a different model based on the orthodox theory of Coulomb blockade\cite{Averin91} was implemented and calculations for similar but isolated dot were performed. The occupancies and conductances for both models showed good qualitative agreement, which points to the Hartree-Fock approach being adequate for describing the Coulomb blockade physics.

The paper is organized as follows. In Sec. II, the Hartree-Fock approach combined with the Green's function formalism is formulated. Sec. III presents the master equation approach to electron tunneling in CB regime. The results of numerical calculations and comparison between two models are given in Sec. IV. Discussion of obtained results is given in Sec. V followed by conclusion.

\section{Model}

\begin{figure}[!tb]
\includegraphics[scale=0.5]{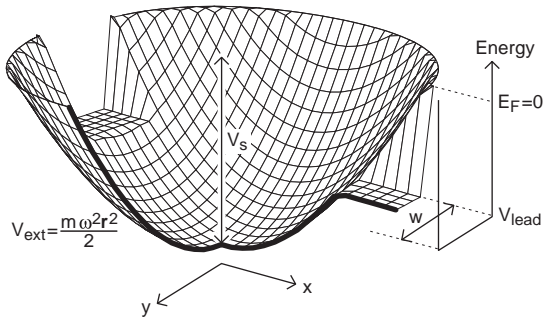}
\caption{External confinement potential $V_{ext}$ of a quantum dot. Only potential below the Fermi energy $E_F=0$ is shown. The parabolic circular potential confines electrons in the cavity attached to two semi-infinite leads of width $w=80$ nm and depth $V_{lead}=-7$ meV. The dot diameter is kept constant, 320 nm, while the number of electron inside the dot is varied. This is accomplished by changing the saddle point potential $V_s$.}
\label{fig:1}
\end{figure}

Consider a two-dimensional open quantum dot attached to semi-infinite leads. Such a system is routinely fabricated in the two-dimensional electron gas by the split-gate technique. The gates induce an external electrostatic confinement $V_{ext}$ that is well described by a harmonic potential, Fig. \ref{fig:1}. 

The Green's function of the system in equilibrium is defined as\cite{Datta_book}
\begin{equation}
 \mathcal{G}^r = \left(E\mathbf{1} + \frac{\hbar^2}{2m^{\ast}}\nabla^2 - V_{ext} - \Sigma^r_{L/R} - \Sigma_H -\Sigma_F \right)^{-1}, 
 \label{eq:gfunction}
\end{equation}
where $m^{\ast}$ is the electron effective mass. The second term is the electron kinetic energy, while last three terms describe effects of the leads\cite{Datta_book} and electron-electron interaction in the Hartree-Fock approximation. The Hartree term accounts for classical electron repulsion
\begin{equation}
\Sigma_H(\mathbf{r})=\frac{e^2}{4\pi \varepsilon _{0}\varepsilon _{r}}\int d%
\mathbf{r}\,^{\prime }n(\mathbf{r}^{\prime })\left( \frac{1}{|\mathbf{r}-%
\mathbf{r}^{\prime }|}-\frac{1}{\sqrt{|\mathbf{r}-\mathbf{r}^{\prime
}|^{2}+4b^{2}}}\right) ,  
 \label{V_H}
\end{equation}
where $n(\mathbf{r})$ is the electron density screened by the mirror charges at the distance of $b=30$ nm from the surface, $\varepsilon_r$ is the dielectric constant, and the integration is performed over the whole device area including the semi-infinite leads. The Fock self-energy can be written in a similar fashion
\begin{equation}
\Sigma_F(\mathbf{r},\mathbf{r}') = -\frac{e^2}{4\pi \varepsilon_0 \varepsilon_r}
n(\mathbf{r},\mathbf{r}')\left( \frac{1}{|\mathbf{r}-%
\mathbf{r}'|}-\frac{1}{\sqrt{|\mathbf{r}-\mathbf{r}'|^2+4b^2}}\right) ,  
 \label{eq:V_F}
\end{equation} 
though its meaning is different: it is the exchange potential that respects quantum mechanical principle of antisymmetry of the wave function. One can show that the Fock term completely compensate the Hartree one for a single electron thus eliminating self-repulsion, see Appendix. In order to remove infinite dimension $\Sigma_F$ is adiabatically damped to zero in the leads. Thus, the dot is treated in the Hartree-Fock approximation but the leads are within the Hartree one. Note that spin degree of freedom is out of scope of present study. The phase relationship among the different states is established by the density matrix
\begin{equation}
 n(\mathbf{r},\mathbf{r}')=-\frac{2}{\pi}\Im \int dE\,\mathcal{G}^r(\mathbf{r},\mathbf{r}',E)\,f_{FD}(E-E_F)
 \label{eq:density}
\end{equation}%
whose diagonal elements are simply the electron density $n(\mathbf{r})$, $f_{FD}(E-E_{F})$ is the Fermi-Dirac distribution function and $E_F$ is the Fermi energy. Eqs. \eqref{eq:gfunction}-\eqref{eq:density} are solved self-consistently, see Ref. \onlinecite{opendot} for details. 

The linear conductance through the quantum dot is given by the Landauer formula\cite{Datta_book} 
\begin{eqnarray}
 &&G = -\frac{2e^2}{h}\int dE\,T(E)\frac{\partial f_{FD}\left( E-E_F\right) }{\partial E}, \\
 &&T(E) = \mathrm{Tr}\left[ \Gamma_L(E)\mathcal{G}^r(E)\Gamma_R(E)\mathcal{G}^a(E)\right]
  \label{eq:conductance}
\end{eqnarray}
where $T(E)$ is the total transmission coefficient, $\mathcal{G}^a(E) = \left[\mathcal{G}^r(E)\right]^{\dag}$ and $\Gamma_{L/R}(E) = 2\Im\left[ \Sigma_{L/R}(E) \right]$ describes coupling the scattering region with the leads. 

\section{Master equation approach}

When the coupling between the dot and leads is weak electron transport is governed by Coulomb blockade (CB) that  in turn is well described by the orthodox theory.\cite{Averin91, Likharev99} It assumes incoherent sequential tunneling of electrons between the dot and electrodes. The approach is generally valid if $G\ll \frac{2e^2}{h}$. Having small coupling between the dot and leads allows one to consider the dot and leads separately and treat the coupling as a perturbation. The Hamiltonian of the system reads
\begin{equation}
  H = H_{dot} + H_{L/R}^{lead} + H^T_{L/R},
  \label{eq:cbhamiltonian}
\end{equation}
where $H_{dot}= \frac{\hbar^2}{2m^{\ast}}\nabla^2 + V_{ext} + V_H$ is the Hamiltonian of the isolated dot and $H_{L/R}^{lead}$ is the Hamiltonian of the isolated left and right leads. The latter is equivalent to lead self-energies in \eqref{eq:gfunction}. The last term in Eq. \eqref{eq:cbhamiltonian} is the standard tunneling Hamiltonian\cite{Averin91} ($L/R$ superscripts for left/right lead omitted here and below otherwise specified) 
\begin{equation}
  H^T = \sum_{i,k} T_{ik} c_i^{\dag}c_k + H.c.
  \label{eq:tunnel}
\end{equation}
with the tunneling matrix elements $T_{ik}$ providing the corresponding tunneling rates
\begin{equation}
  \Gamma_k(E) = \frac{2\pi}{\hbar} \sum_i \left| T_{ik} \right|^2 \delta(E-E_i).
  \label{eq:rates}
\end{equation}
The total rates $\Gamma^{\pm}$ of electron tunneling to/from the dot are the sum of all partial tunneling rates to/from specific energy levels $E_k$
\begin{eqnarray}
  \Gamma^+(N) &=& \sum_k \Gamma_k(E) f_{FD}(E-E_F) \left[ 1-g_N(E_k) \right], \\
  \Gamma^-(N) &=& \sum_k \Gamma_k(E) \left[ 1-f_{FD}(E-E_F) \right] g_N(E_k),
  \label{eq:totrates}
\end{eqnarray}
where $g_N(E_k)$ is the single particle distribution function associated with the Gibbs distribution $F(E_{k_1},\ldots,E_{k_N})$ of $N$ electrons in the dot,
\begin{eqnarray}
  &&g_N(E_k) = \sum_{k_1,\ldots,k_N} F(E_{k_1},\ldots,E_{k_N}), \label{eq:gibbs1} \\
  &&F(E_{k_1},\ldots,E_{k_N}) = \frac{1}{Z} exp \left[ -\frac{1}{k_BT}\left( \sum_i E_{k_i} -  \mu_N N \right) \right], \\ \nonumber
  &&Z = \sum_{k_1,\ldots,k_N} exp \left[ -\frac{1}{k_BT}\left( \sum_i E_{k_i} -  \mu_N N \right) \right]. \nonumber
  \label{eq:gibbs2}
\end{eqnarray}
Here $Z$ is a normalization constant, and $\mu_N$ denotes the chemical potential in the dot occupied by $N$ electrons. To calculate tunneling rates $\Gamma^{\pm}$, the self-consistent solution for $H_{dot}\Psi=E\Psi$ should be found first. Because we are not interested in CB details but rather need to test predictions of the Hartree-Fock model established in previous section the electron-electron interaction in the dot is restricted to the Hartree approximation; $V_H$ is given by Eq. \eqref{V_H} with integral restricted by the dot area. 

The electron transport in the Coulomb blockade regime is described by a kinetic ``master'' equation for $p(N)$, the probability that there are $N$ electrons in the dot\cite{Likharev99, Averin91}
\begin{eqnarray}
	\Gamma^+(N-1) p(N-1) + \Gamma^-(N+1) p(N+1) = \\ \nonumber
	\left[ \Gamma^-(N) + \Gamma^+(N) \right] p(N).
	\label{eq:master}
\end{eqnarray}
The average current for any of the tunneling junctions is calculated as (with $L/R$ superscripts explicitly written)
\begin{equation}
	I_{L/R} = e \sum_N p(N)\left[ \Gamma^+_{L/R}(N) - \Gamma^-_{L/R}(N) \right].
\end{equation} 
This readily allows one to obtain the conductance $G=I_LV=-I_RV$.

\begin{figure*}[tb]
\includegraphics[keepaspectratio,width=\textwidth]{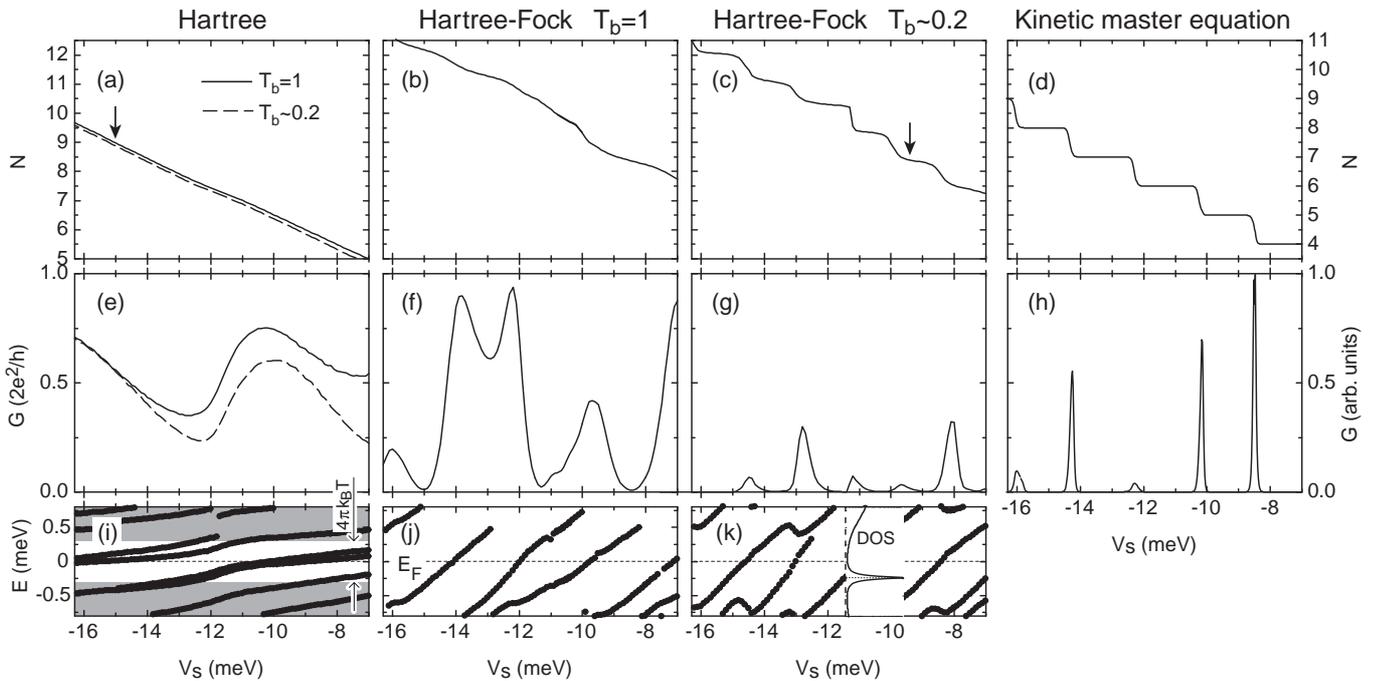}
\caption{Occupancy $N$ (a)-(d), conductance $G$ (e)-(h) and energy structure (i)-(k) of the dot calculated in the Hartree, Hartree-Fock, and master equation approaches. The Hartree and Hartree-Fock results in panels (a), (e), (i) and (c),(g),(k), respectively, corresponds to potential barriers of transmittance $T_b\sim0.2$ inserted at entrance and exit of the dot. Insert in (k) shows the DOS as a function of energy that identify resonant level position. Arrows in (a) and (c) mark $V_s$ potential for charge density plots in Fig. \ref{fig:3}. Temperature 0.5 K.}
\label{fig:2}
\end{figure*}

\begin{figure}[tb]
\includegraphics[keepaspectratio,width=\columnwidth]{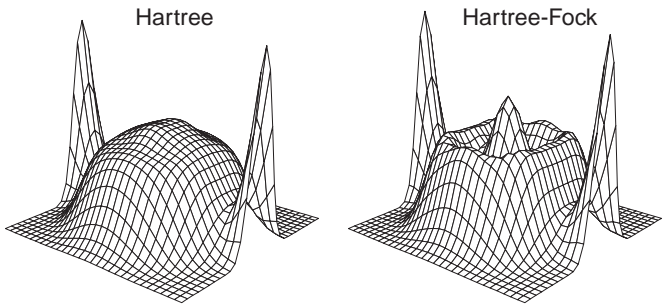}
\caption{Representative charge densities in the dot calculated in the Hartree and Hartree-Fock approaches for similar dot's occupancy, see arrows in Fig. 2(a),(c).}
\label{fig:3}
\end{figure}

\section{Results}

In order to test predictions of the Hartree-Fock approach we compare them with the results obtained using the Hartree and master equation approaches. The Hartree approximation is known to be correct for open systems where several propagating states enter and leave the scattering region, i.e. $G>\frac{2e^2}{h}$. The master equation is, in opposite, well suitable for description of closed systems where CB physics dominates, i.e. $G\ll\frac{2e^2}{h}$.\cite{Likharev99} As a testing platform we consider an open quantum dot with one propagating mode available for electrons to enter and leave the dot, Fig. \ref{fig:1}, and then will impose the tunneling barriers at the dot openings. The barriers are accomplished by a potential raise and equivalent to the quantum point contact constrictions formed in experimental setups.

As a model system we consider an open quantum dot defined by split-gates at GaAs-AlGaAs heterointerface. Electron band sturcture in GaAs is well described by parabolic dispersion with $m^{\ast}=0.067m$; $\varepsilon_r=12.9$. Note however that the model is not material specific and might also be applied to other two-dimensional structures, e.g. Si inversion layer structures. The parameters of the confinement potential are indicated in the Figure \ref{fig:1}. 

The Figs. \ref{fig:2}(a),(e),(i) show the number of electrons $N$, conductance and energy structure as a function of the saddle point potential $V_s$ calculated in the Hartree approximation. Its remarkable features are linear change of $N$ and energy level pinning effect.\cite{opendot, Hirose, dotPRL} The latter is pronounced as the resonant levels track the Fermi energy within $\pm 2\pi k_B T$ interval, Fig. \ref{fig:2}(i). Each level represents the enhanced density of states (DOS) broadened by coupling to the extended states in the leads. By minimizing the total energy those peaks shift causing metallic-like behavior of electrons in the dot. The screening ability of the dot is greatly enhanced allowing electrons to easily change their location in respect to external electric field. One may loosely estimate the energy level separation as $\Delta=0.5$ meV, Fig. \ref{fig:2}(i). The conductance through the dot depends on coupling of particular resonant states with the extended states in the leads.\cite{opendot, Zozoulenko_1997} Because other states in the dot couple less effectively they sweep through the Fermi energy causing no effect of the conductance but contributing to the occupancy. This explains the period broadening of the conductance oscillations in the Hartree approximation.\cite{opendot}

If the Fock term is taken into account the conductance and energy structure change dramatically, Figs. \ref{fig:2}(b),(f),(j). The energy structure becomes more complicated with levels near $E_F$ ordered and less pinned. The level separation $\Delta$ about doubles in comparison to the Hartree approach. The conductance oscillations become narrower. Each conductance peak corresponds to the resonant level crossing the Fermi energy. The distance between successive peaks equals roughly to the potential $V_s$ needed to change dot occupancy by one electron. However, $N$ still follows linear dependence on $V_s$. It becomes step-like if the potential barriers separating the dot from the leads are imposed. The transmission of each barrier equals to $T_b\sim0.2$ in $V_s$ range of Figs. \ref{fig:2}(c),(g),(k). The conductance for a weakly coupled dot generally decreases and well separated peaks become clearly visible. Inspection of the electron density reveals strong Freidel-like oscillations caused by the exchange interaction in the Hartre-Fock approach, see Fig. \ref{fig:3}.

The Figs. \ref{fig:2}(d),(g) show the results of the master equation approach for the dot of the same geometry as studied above. The dot however is isolated from the leads and treated separately. The agreement between master equation and Hartree-Fock approximation for weakly coupled dot is remarkable: the electron number is quantized and conductance shows periodic peaks. Each peak occurs when occupancy changes by one, i.e. in between of quantized $N$ plateau. All these features are manifestation of the CB physics.

\section{Discussion}

Incorporation of the electron-electron interaction in the Hartree approximation proved to be sufficient if coupling between scattering region and leads is strong,\cite{dotPRL,Ihnatsenka,agreeOthers} $G>\frac{2e^2}{h}$. It was shown that Hartree and DFT approaches provide qualitatively similar description of the electron transport in open quantum dots.\cite{opendot} Using the fact that DFT predicts spin polarization in a quantum wire similarly to the Hartree-Fock approximation\cite{JPCM} we may conclude that all these approaches are qualitatively equivalent for description of electron transport in the strong coupling regime $G\gg\frac{2e^2}{h}$. However, in the opposite regime of weak coupling, the Hartree approximation is not accurate. That regime is governed by Coulomb blockade physics when the electron density is quantized and therefore the Hartree-Fock approach should be rather employed. Incorporating the density matrix into the theory is essential because no features of charge quantization and CB are found otherwise. Even though the exchange interactions might be treated within DFT in the local density approximation or Slater approximation\cite{Giuliani_Vignale} they fail to address charge quantization and CB physics.

The important feature of the present model is absence of any adjustable parameters. In this respect, it might be considered as a first-principle approach. The models existing in literature use adjustable charging parameters and/or disregard degrees of freedom of the scattering region,\cite{Henrickson, Beenakker91, Jovanovic, Palacios05, Averin91, Meir91, Wacker} e.g. the dot is treated as a zero-dimensional region.\cite{Beenakker91, Meir91} Such simplifications make it difficult to relate the employed parameter set with physical processes they represent and, as a result, the reliability of the results obtained is questionable. Note that some studies consider the device geometry explicitly but still rely on some assumption heavily affecting the obtained results. For example, slight modification of the Landauer formula allowed authors of Ref. \onlinecite{Jovanovic} to address CB even though amplitude of successive conductance peaks fell off exponentially, which disagree with experimental fundings. 

While the Hartree-Fock model appears to be powerful for the description of electron transport it is still a mean-field approximation to an interacting many-electron system. The model misses correlation effects that stem from the fact that true ground state wave function is not a single Slater determinant.\cite{Giuliani_Vignale} Electron correlation interactions might play an important role as it is shown for the Kondo effect and others.\cite{Kurth10, Feretti05} It worth noting also that the Hartree-Fock model is known to underestimate electron screening.\cite{Giuliani_Vignale} A possible improvement might be accomplished by treating electrons participating in transport by some exact method while using the mean-field approach for others.

Numerical calculations in the Hartree-Fock approach demand much larger computer resources in comparison to the Hartree approximation. The matrix dimensions scale as $(n\times m)^2$ and $n^2\times m$ for these approaches, respectively ($n$ and $m$ are the numbers of sites in transverse and longitudinal directions, respectively). In practice, the transverse space coordinate is transformed into momentum space where several basis functions are used.\cite{Ihnatsenka} Convergence of the self-consistent solution in the Hartree-Fock approach performs generally worse, especially when lead-to-dot coupling decreases substantially and $G\ll\frac{2e^2}{h}$. A possible improvement might be done as suggested in Ref. \onlinecite{Indlekofer05}, where the Fermi-Dirac distribution function in Eq. \eqref{eq:density} is replaced by the Gibbs distribution. 

Finally, it worth noting that the similar results were obtained for the open quantum dots of different geometries. The numerical calculations were also performed for the quantum point contacts, where conductance plateaus were found to be somewhat larger if the Hartree-Fock approach employed. This signals reduced energy level pinning effect in accord with results presented above. 

\section{Conclusion} 

The Hartree-Fock approximation combined with the Green's function formalism represents an unified and powerful approach to electron transport that implicitly takes the interplay between the statistical and quantum-mechanical properties of the confined geometries into account. This allows one to address the Coulomb blockade physics that dominates electron transport in quantum dots weakly coupled to the leads. This is confirmed by comparison with the master equation approach for equivalent closed dots: the number of electrons inside the dot is quantized and the conductance shows a peak each time the electron number changes by one (disregarding spin). The conductance peaks are caused by electron resonant transmission when corresponding energy levels cross the Fermi energy. 

The present study is limited to the case of spinless electrons. Work is in progress to include the effect of the spin in order to revisit magnetic bound state formation in the quantum point contacts. It is also interesting to examine multiple periodicity in the Aharonov-Bohm interferometers as well as the effect of mesoscopic CB observed in the recent experiment\cite{Amasha10} on the open quantum dot.

\bigskip

\begin{acknowledgments}
I am grateful to I.V. Zozoulenko and G. Kirczenow for discussions and critical reading of the manuscript. 
\end{acknowledgments}

\appendix

\section{Appendix: Self-interaction correction in the Hartree-Fock approach}

In order to understand why the Hartree-Fock approach eliminates self-interaction error, let us consider a system of $N$ electrons each described by the Hamiltonian $h_i$. The ground state energy reads
\begin{equation}
E = \sum_{i\sigma}^N \left\langle i\left|h_i\right|i\right\rangle + \sum_{i\sigma}^N \sum_{j>i\:\sigma^{\prime}}^N \left( J_{i\sigma ; j\sigma^{\prime}} - K_{i\sigma ; j\sigma^{\prime}} \right),
  \label{eq:SIC1}
\end{equation}
where $J$ represents Coulomb interaction between electrons in state $i$ and $j$
\begin{equation}
J_{i\sigma ; j\sigma^{\prime}} = \int \int \frac{\left| \Psi_{i\sigma}(\mathbf{r}_1) \right|^2 \left| \Psi_{j\sigma^{\prime}}(\mathbf{r}_2) \right|^2}{\left| \mathbf{r}_1-\mathbf{r}_2\right|} d\mathbf{r}_1 d\mathbf{r}_2
  \label{eq:SIC2}
\end{equation}
and $K$ represents the effect of antisymmetrization, i.e. exchange,
\begin{equation}
K_{i\sigma ; j\sigma^{\prime}} = \int \int \frac{ \Psi_{i\sigma}^*(\mathbf{r}_1) \Psi_{i\sigma^{\prime}}^*(\mathbf{r}_2) \Psi_{j\sigma^{\prime}}(\textbf{r}_2) \Psi_{i\sigma}(\textbf{r}_1) }{\left| \mathbf{r}_1-\mathbf{r}_2\right|} d\mathbf{r}_1 d\mathbf{r}_2
  \label{eq:SIC3}
\end{equation}
Since $J_{i\sigma ; i\sigma^{\prime}}=K_{i\sigma ; i\sigma^{\prime}}$ electron in state $i$ does not self interact. This holds true regardless the form of the wave function. Note that in DFT, the Coulomb term $J$ is the same as in the Hartree-Fock but the exchange term is approximate. So cancelation of self-interaction in Coulomb and exchange terms is incomplete.

\end{document}